\newcommand{\vct}[1]{{\bf #1}}
\newcommand{\eq}[1]{Eq.~(\ref{eq:#1})}

\documentclass[preprint,prb,showpacs,preprintnumbers,amsmath,amssymb]{revtex4}

\usepackage{graphicx}
\usepackage{dcolumn}
\usepackage{bm}
\usepackage[usenames]{color}

\begin{document}


\title{Zero-point Divacancy Concentration in the Shadow Wave-Function Model for Solid $^4$He}

\author{R. Pessoa}
\author{M. de Koning}
\author{S. A. Vitiello}
\affiliation{Instituto de F\'{\i}sica Gleb Wataghin, Caixa Postal 6165 \\
Universidade Estadual de Campinas - UNICAMP \\
13083-970, Campinas, SP, Brazil}

\date{\today}

\begin{abstract}
We address the issue of interaction between zero-point vacancies in solid $^4$He as described within the shadow wave-function model.
Applying the reversible-work method and taking into account finite-size effects, we obtain a zero-point monovacancy concentration
of $(2.03\pm 0.02)10^{-3}$, which is slightly higher than the result due to Reatto {\em et al.} for the same model. Utilizing the same methodology,
we then consider the divacancy, taking into account both the in-plane as well as out-of-plane configurations with respect to the basal plane. We find no significant anisotropy between both conformation. Furthermore, although there is a small binding tendency, the expected divacancy concentration is only $\sim 4-5$ times larger than the value expected in the absence of any clustering propensity, $2.5 \times 10^{-5}$. This result suggests that, within the employed model description, no vacancy aggregation leading to phase separation is to be expected in the ground state.
\end{abstract}


\maketitle


The experimental observation of a non-classical rotational inertia (NCRI) by Kim and Chan~\cite{Kim2004,Kim2004a} in solid $^4$He and its interpretation as a possible manifestation of mass superflow in the solid phase, has triggered intensive research efforts both at experimental and theoretical levels~\cite{Prokof'ev2007,Balibar2008,Galli2008}. While there is still no consensus as to the precise origin of the observed phenomenology, it is a generally accepted notion that it is not an intrinsic property of the pristine, defect-free crystalline phase, but rather that it is somehow related to crystal disorder, in the form of crystal defects such as vacancies and interstitials, dislocations and grain boundaries~\cite{Anderson2005,Pollet2007,Boninsegni2007,Day2007,Rossi2008,Pollet2008,Corboz2008,Day2009,Anderson2009} and/or the presence of glassy regions~\cite{Saunders2009,Hunt2009}.

In this context, the role of vacancies has received a significant amount of attention ever since the first theoretical proposals of a supersolid ground state. Yet, their role remains controversial to this day. On the one hand, finite-temperature path-integral Monte Carlo (PIMC) calculations seem to suggest that the ground state of solid $^4$He does not contain vacancies~\cite{Boninsegni2006,Clark2006}. On the other hand it has been argued that a number of technical issues, including the use of periodic boundary conditions and small numbers of particles in the simulation box, may in fact prevent such calculations from correctly assessing the true nature of the zero-temperature ground state~\cite{Galli2006,Rossi2008}. Moreover, experimental data indicating that a vacancy concentration below 0.4\% cannot be ruled out,~\cite{Simmons2007} as well as arguments due to Anderson and co-workers~\cite{Anderson2005,Anderson2009}, contend the possibility of zero-point vacancies in solid $^4$He. In this light the question whether or not solid $^4$He contains a finite zero-point vacancy concentration remains open.

In case of the existence of a finite zero-point vacancy concentration, an issue of concern involves the interaction between vacancies~\cite{Mahan2006,Rossi2008} and a possible propensity toward clustering. Two recent studies~\cite{Mahan2006,Rossi2008} have addressed this issue from two different points of view. Mahan and Shin performed a theoretical study of the interaction between two fixed vacancies in solid $^4$He using elasticity theory of the $hcp$ crystal. Their results indicate that the interaction is attractive and anisotropic: the divacancy interaction energy was found to be more strongly attractive along the $c$-axis compared to directions within the basal plane. These calculations, however, do not explicitly include the effects associated with the zero-point motion of the vacancies. Rossi and coworkers~\cite{Rossi2008} addressed this issue by performing calculations based on the shadow wave function (SWF) description~\cite{Vitiello1988,MacFarland1994} which has bee
 n able to reproduce many of the properties of $^4$He in the solid phases~\cite{MacFarland1994,Moroni1998}. Their approach is based on the well-known equivalence between the calculation of the zero-point concentration of point defects in the ground state of a bosonic quantum system and that of an associated classical solid~\cite{Hodgdon1995,Rossi2008,Pessoa2009} at a finite temperature. Employing this approach and the thermodynamic integration (TI) technique~\cite{Frenkel2002}, in which the formation free energy of the vacancy is determined by subtracting the free energies of the computational cells with and without vacancy, they determine the zero-point monovacancy concentration within the SWF model. The influence of periodic-image effects, however, due to the elastic interaction between the periodic images of the vacancy, were not taken into account. In addition to the monovacancy, vacancy-vacancy interactions were studied by observing systems containing two and three vacan
 cies and collecting statistics in terms of a vacancy-vacancy correlation function. However, the actual zero-point divacancy concentration was not determined.

In this Brief Report we refine the studies carried out by Rossi and coworkers for the SWF model of solid $^4$He in three ways. First, we provide a more accurate result for the zero-point monovacancy concentration, taking into account elastic interactions between periodic images. To this end we use the reversible-work (RW) method~\cite{deKoning2004}, which allows a direct computation of the work required to reversibly introduce a vacancy in an initially defect-free crystal without the need for the subtraction of large numbers, and a finite-size extrapolation. This approach was recently applied in the context of bosonic quantum crystals to determine the zero-point vacancy concentration of a system described by the Jastrow wave function~\cite{Pessoa2009}. Secondly, applying the same computational scheme, we determine the zero-point concentration of the divacancy for the SWF model. Finally, to detect a possible anisotropy as in Ref.~\onlinecite{Mahan2006} we distinguish between the in-plane and out-of-plane divacancy configurations with respect to the basal plane.


Formally, the true wave function of a system of $N$ bosons in its ground state can be written as~\cite{Hodgdon1995}
\begin{equation}
\Psi(R) = \exp\left[- \frac{1}{2} \Phi(R)\right] / Q_N^{1/2}\; , \label{eq:wf}
\end{equation}
where $R \equiv \left\lbrace {\bf r}_{1},{\bf r}_{2}, \ldots, {\bf r}_{N} \right\rbrace $
stands for the particle coordinates, $\Phi$ is an effective potential, and $Q_N$ is a normalization constant. Since $\psi(R)$ is positive everywhere, $\vert\psi(R)\vert^2$ can, without loss of generality, be interpreted as a Boltzmann factor of a classical system described by the potential function $\Phi(R)$ at a temperature $k_B T=1$.
In the SWF variational theory of $^4$He,~\cite{MacFarland1994} the ground-state wave function is given as
\begin{equation}
 \Psi (R) = \frac{1}{Q^{1/2}_N}\psi_{J}(R) \int dS \prod_{i}\theta({\bf r}_{i}-{\bf s}_{i}) \psi_{S}(S) \label{eq:Psi},
\end{equation}
where the $\psi_J$ and $\psi_{S}$ are the Jastrow factors
\begin{equation}
\psi_{J}(R) = \exp\left[ -\frac{1}{2} \sum_{i<j}^{N}\left( \frac{b}{r_{ij}}\right)^{5} \right]
\end{equation}
with $r_{ij} = |\vct r_i - \vct r_j|$, and
\begin{equation}
\psi_{S}(S) = \exp\left[ - \sum_{i<j}^{N} \delta V(\alpha s_{ij}) \right].
\end{equation}
The latter depends on a rescaled atomic potential $V$, here taken to be the Aziz potential,~\cite{Aziz1979} and on the distance $s_{ij} = |\vct s_i - \vct s_j|$ between auxiliary variables $i$ and $j$ of the set $S \equiv \{\vct s_1, \vct s_2, \ldots, \vct s_N\}$. This set of variables is coupled to the particles coordinates $R$ through a Gaussian factor, $\theta({\bf r}_{i}-{\bf s}_{i}) = \exp\left[ -C \vert {\bf r}_{i} - {\bf s}_{i} \vert ^{2}\right] $. The volume integration in $dS = d\vct s_1 d\vct s_2 \cdots d\vct s_N$ is over the whole space. The variational parameters  $b$, $C$, $\delta$ and $\alpha$ are those that minimize the expectation value of the energy.~\cite{MacFarland1994}

Within this formulation, the quantum-mechanical probability-density function $|\Psi(R)|^2$ can be associated with the Boltzmann factor of the classical system described by the effective potential
\begin{eqnarray}
\nonumber
 \Phi(R,S,S') &=&  \sum_{i<j}^{N}
\left ( \frac{b}{r_{ij}}  \right )^5
+ C\sum_{i}^{N} \left [ \vert {\bf r}_{i}-{\bf s}_{i} \vert^{2} + \vert {\bf r}_{i}-{\bf s}'_{i} \vert^{2} \right ] \\
&+& \delta\sum_{i<j}^{N} \left [ V(\alpha s_{ij})+  V(\alpha s'_{ij}) \right ] .
 \label{eq:veff}
\end{eqnarray}
This fictitious system can be thought as composed of $N$ interacting trimers, each one composed of one actual atom and a pair of coupled shadow degrees of freedom. Given the aforementioned equivalence between the quantum system and this classical system, the zero-point vacancy and divacancy concentrations in the quantum system described by \eq{Psi} is then equal to the thermal equilibrium vacancy and divacancy concentrations in the system defined by \eq{veff} at a temperature $k_B T=1$.

In order to determine these concentrations we now follow the RW method,~\cite{deKoning2004,Pessoa2009} which allows the computation of the formation free energies of the respective defect configurations in the fictitious classical system. The RW approach is based on the construction of continuous thermodynamic paths that connect a system of interest to a certain reference. In practice this is achieved by introducing one or more coupling parameters that measure the progress along the given path. By measuring the reversible work along this path one can obtain the free-energy difference between the system of interest and the reference. In the particular cases of the thermal equilibrium vacancy and divacancy concentrations, we are interested in the formation free energies
\begin{eqnarray}
\nonumber
\Delta F_{\rm m} &\equiv& F(N-1)-\frac{N-1}{N} F(N) \\
&=& \frac{1}{N} F(N)+[F(N-1)-F(N)]
\label{eq:deltaF}
\end{eqnarray}
and
\begin{eqnarray}
\nonumber
\Delta F_{\rm d} &\equiv& F(N-2)-\frac{N-2}{N} F(N) \\
&=& \frac{2}{N} F(N)+[F(N-2)-F(N)],
\label{eq:deltaF2}
\end{eqnarray}
respectively. Here $F(N)$ represents the free energy of a defect-free crystal containing $N$ atoms, $F(N-1)$ is the free energy of a crystal containing $N-1$ atoms and a monovacancy and $F(N-2)$ is the free energy of a crystal containing $N-2$ atoms and two vacancies adjacent to each other. According to the second lines of \eq{deltaF} and \eq{deltaF2} they can be written in terms of the free energy per atom of the defect-free system and the free-energy differences between a monovacancy, (divacancy), cell containing $N-1$, $(N-2)$, atoms and the defect-free cell with $N$ atoms. Using this partition, we construct separate thermodynamic paths that allow us to compute both contributions. As detailed in Refs.~\onlinecite{deKoning2004}~and~\onlinecite{Pessoa2009} the reversible work values along these paths are measured along finite-time nonequilibrium simulations during which the coupling parameters are varied dynamically. In order to eliminate the systematic errors associated wit
 h the nonequilibrium nature of these simulations, the switching processes are carried out in both directions.

To determine the free-energy differences in Eqs.~\eq{deltaF} and \eq{deltaF2} we employ thermodynamic paths that involves a continuous transformation of the defect-free interacting fictitious classical system such that one or more of its atoms are decoupled from the remainder of the system. At the same time these atoms are transformed into a collection of noninteracting classical harmonic oscillators. Given the specific functional form of the effective classical potential in case of the SWF model, we define the thermodynamic path
\begin{eqnarray}
\nonumber
 &U&({\bf R},{\bf S},{\bf S}';\{\lambda\}) = \sum_{i<j}^{N} \lambda_i\,\lambda_j
\left ( \frac{b}{r_{ij}}  \right )^5 \\
\nonumber
&+& C\sum_{i}^{N} \lambda_i \left [ \vert {\bf r}_{i}-{\bf s}_{i} \vert^{2} + \vert {\bf r}_{i}-{\bf s}'_{i} \vert^{2} \right ]  \\
\nonumber
&+& \delta\sum_{i<j}^{N} \lambda_i\lambda_j\left [ V(\alpha s_{ij})+  V(\alpha s'_{ij}) \right ] +\frac{1}{2}\sum_{i=1}^{N}(1-\lambda_i) \\
&\times& \left( \kappa_1\vert {\bf r}_{i}-{\bf r}_{i}^{(0)} \vert^2 +  \kappa_2\vert {\bf s}_{i}-{\bf r}_{i}^{(0)} \vert^2 + \kappa_2\vert {\bf s}'_{i}-{\bf r}_{i}^{(0)} \vert^2\right)
\label{eq:SWpath}
\end{eqnarray}
in which each trimer $i$ in the system is coupled to a switching parameter $\lambda_i$, that is allowed to vary between 0 and 1. When $\lambda_i=1$ for all $N$ particles $i$, the system corresponds to the fictitious system described by the effective potential \eq{veff}. Similarly, when $\lambda_i=0$ for all $i$ the system corresponds to a collection of $3N$ noninteracting harmonic oscillators. In this case, in addition to the degrees of freedom of the atom ${\bf r}_{i}$, the coordinates ${\bf s}_{i}$ and ${\bf s}'_{i}$ describing the auxiliary degrees of freedom are also transformed into harmonic oscillators, centered about the lattice sites ${\bf r}_{i}^{(0)}$. The spring constants $\kappa_1$ and $\kappa_2$ are chosen such that the mean-square displacement of the harmonic oscillators is approximately equal to that of the atoms and auxiliary degrees of freedom in the fully interacting system.~\cite{Frenkel2002,Pessoa2009}

Using the general path defined by ~\eq{SWpath} we can determine both contributions in~\eq{deltaF} and ~\eq{deltaF2}. To compute $F(N)$, we set $\lambda_i=\lambda$ for all $i$, with $\lambda$ varying between 0 and 1. The corresponding thermodynamic path describes a transformation from the defect-free interacting system into a collection of $3N$ noninteracting classical harmonic oscillators, for which the free energy is known analytically. To compute $F(N-1)-F(N)$, we fix $\lambda_i=1$ for $i\neq k$, and switch $\lambda_k$ between 0 and 1.  In this manner, the trimer associated with particle $k$ is decoupled from the remainder of the system, turning the defect-free interacting system into a system containing a monovacancy at lattice site $k$ plus 3 independent harmonic oscillators. In a similar fashion, by fixing  $\lambda_i=1$ for $i\neq k,l$ and varying $\lambda_k$ and $\lambda_l$ between 0 and 1, one inserts two vacancies into the defect-free interacting system, plus 6 independent harmonic oscillators. When the lattice sites associated with atoms $k$ and $l$ are nearest neighbors in the lattice, this corresponds to the creation of a divacancy. As usual, to avoid singularities in the calculation of the reversible work, the vacancies are not allowed to diffuse during the reversible creation process, restricting the motion of their nearest-neighbor particles to their respective Wigner-Seitz primitive cells.~\cite{Pessoa2009,deKoning2004} For convenience the center of mass of the system is held fixed during the switching simulations. The volume during the switching simulations is held fixed. In order for the results to become independent of this imposed boundary condition, a finite-size extrapolation is required.

In our calculations we apply the Metropolis algorithm to sample configurations from $\vert \Psi \vert^2$. We use orthorhombic simulation cells with a $hcp$ structure and numbers of atoms varying between $180$ and $700$ at the melting density $\rho = 0.0294 \textrm{\AA}^{-3}$ subject to periodic boundary conditions. Before starting each switching process the system is equilibrated for at least $2\times10^4$ Monte Carlo (MC) sweeps where $3N$ random attempts are made to move an atom or shadow variable. All switching processes are performed using $1.5\times10^4$ MC sweeps per process, which is sufficiently slow to guarantee the regime of linear response. The estimates of reversible work are obtained as averages over $120$ independent forward and backward switching processes.

\begin{figure}
\includegraphics*[scale=0.5]{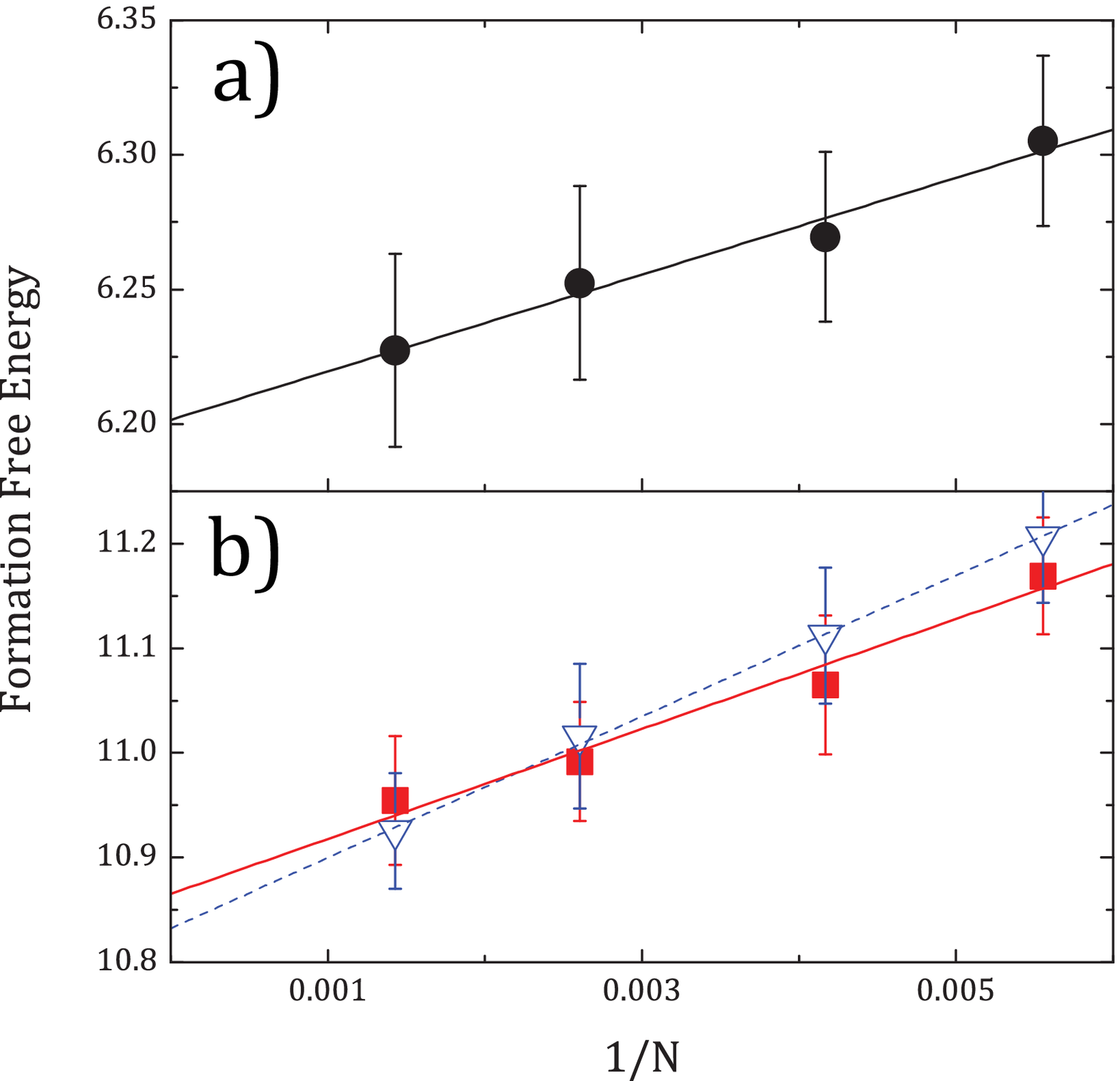}
\caption{\label{Figure1} (Color online) Vacancy formation free energies as a function of inverse particle number in the computational cell. Free energies are measured in units of $k_B T=1$. a) Monovacancy. Symbols with error bars represent RW data. Line denotes linear regression. b) Divacancy. Triangles represent RW data obtained for divacancy perpendicular to basal plane. Square denote RW data for divacancy in the basal plane. Dashed and full lines denotes the respective linear regressions.}
\end{figure}

The results of the calculations are shown in~Fig.~\ref{Figure1}, which shows the monovacancy and divacancy formation free energies as a function of the inverse particle number in the computational cell. The monovacancy formation free energy is seen to slightly decrease with increasing system size, which is a consequence of the elastic image interactions due to the periodic boundary conditions. The same occurs for the divacancy formation free energies as depicted in Panel b). By plotting the results as a function of $1/N$ and extrapolating to $N\rightarrow \infty$ by means of linear regressions, we find estimates for the isolated mono and divacancy formation free energies. For the monovacancy we obtain $\Delta F_{\rm m} = 6.20 \pm 0.01 $, in units of $k_B T$=1, which gives rise to an equilibrium concentration of $c_m = \exp(-\Delta F_{\rm m})=(2.03 \pm 0.02)\times 10^{-3}$. This result is slightly higher than the value $c_m = (1.4 \pm 0.1)\times 10^{-3}$ reported by Rossi {\em
et al.}~\cite{Rossi2008}. The results for the divacancy suggest that, in contrast to the findings in Ref.~\onlinecite{Mahan2006}, there is no significant anisotropy with respect to the basal plane. The extrapolated formation free energies are $\Delta F_{\rm d} = 10.87 \pm 0.06 $ and $10.84 \pm 0.06 $ for the in-plane and out-of-plane divacancy configurations, respectively. Furthermore, comparison with the extrapolated monovacancy formation free energy $\Delta F_{\rm m}$ shows a positive binding free energy $F_{\rm b}=2F_v-F_{2v}$ for the divacancy, indicating a driving force toward clustering. However, despite the positive value of the binding free energy, the corresponding clustering tendency is found to be rather modest. The divacancy concentration, given by $c_{d}=(z/2) c_m^2 \exp(F_b)$ where $z$ is the number of nearest-neighbor sites in the lattice, is only 4-5 times larger than the divacancy concentration in the case of a zero binding energy, $2.5 \times 10^{-5}$.
This suggests that, although two vacancies do form a bound state, the binding tendency within the present model is not sufficiently large to provoke vacancy aggregation leading to large-scale phase separation. This is in agreement with the observation of Rossi {\em et al.}~\cite{Rossi2008}.

In summary, we address the issue of interaction between zero-point vacancies in solid $^4$He as described within the shadow wave-function model.
Using the reversible-work method taking into account finite-size effects, we consider the for both the in-plane and out-of-plane configurations with respect to the basal plane. In addition to finding no significant anisotropy between both conformation, the expected divacancy concentration is only $\sim 4-5$ times larger than the value expected in the absence of any clustering propensity. These results suggest that, within the employed model description, no vacancy aggregation leading to phase separation is to be expected in the ground state.

The authors acknowledge financial support from the Brazilian agencies FAPESP, CNPq and CAPES. Part of the computations were performed at the CENAPAD high-performance computing facility at Universidade Estadual de Campinas.


\pagebreak

\end{document}